# Continuum Percolation in the Intrinsically Secure Communications Graph

Pedro C. Pinto, *Student Member, IEEE*, and Moe Z. Win, *Fellow, IEEE*

*Abstract*—The intrinsically secure communications graph ($i\mathcal{S}$-graph) is a random graph which captures the connections that can be securely established over a large-scale network, in the presence of eavesdroppers. It is based on principles of information-theoretic security, widely accepted as the strictest notion of security. In this paper, we are interested in characterizing the global properties of the $i\mathcal{S}$-graph in terms of percolation on the infinite plane. We prove the existence of a phase transition in the Poisson $i\mathcal{S}$-graph, whereby an unbounded component of securely connected nodes suddenly arises as we increase the density of legitimate nodes. Our work shows that long-range communication in a wireless network is still possible when a secrecy constraint is present.

## I. INTRODUCTION

Percolation theory studies the existence of phase transitions in random graphs, whereby an infinite cluster of connected nodes suddenly arises as some system parameter is varied. Percolation theory has been used to study connectivity of multi-hop wireless networks, where the formation of an unbounded cluster is desirable for communication over arbitrarily long distances.

Various percolation models have been considered in the literature. The Poisson Boolean model was introduced in [1], which derived the first bounds on the critical density, and started the field of continuum percolation. The Poisson random connection model was introduced and analyzed in [2]. The Poisson nearest neighbour model was considered in [3]. The signal-to-interference-plus-noise ratio (SINR) model was characterized in [4]. A comprehensive study of the various models and results in continuum percolation can be found in [5]. Secrecy graphs were introduced in [6] from an information-theoretic perspective, and in [7] from a geometrical perspective. The local connectivity of secrecy graphs was extensively characterized in [8], while the scaling laws of the secrecy capacity were presented in [9]. The effect of eavesdropper collusion on the secrecy properties was studied in [10].

In this paper, we characterize long-range secure connectivity in wireless networks by considering the *intrinsically secure communications graph* ($i\mathcal{S}$-graph), defined in [8]. The $i\mathcal{S}$-graph describes the connections that can be established with information-theoretic security over a large-scale network. We prove the existence of a phase transition in the Poisson $i\mathcal{S}$-graph, whereby an unbounded component of securely connected nodes suddenly arises as we increase the density of legitimate nodes.

P. C. Pinto and M. Z. Win are with the Laboratory for Information and Decision Systems (LIDS), Massachusetts Institute of Technology, Room 32-D674, 77 Massachusetts Avenue, Cambridge, MA 02139, USA (e-mail: ppinto@alum.mit.edu, moewin@mit.edu).

This research was supported, in part, by the MIT Institute for Soldier Nanotechnologies, the Office of Naval Research Presidential Early Career Award for Scientists and Engineers (PECASE) N00014-09-1-0435, and the National Science Foundation under grant ECCS-0901034.

In particular, we determine for which combinations of system parameters does percolation occur. Our work shows that long-range communication in a wireless network is still possible when a secrecy constraint is present.

This paper is organized as follows. Section II describes the system model. Section III introduces various definitions. Section IV presents the main theorem, whose underlying lemmas are proved in Sections V and VI. Section VII provides additional insights through numerical results. Section VIII presents some concluding remarks.

## II. SYSTEM MODEL

### A. Wireless Propagation Characteristics

Given a transmitter node $x_i \in \mathbb{R}^d$ and a receiver node $x_j \in \mathbb{R}^d$, we model the received power $P_{\mathrm{rx}}(x_i, x_j)$ associated with the wireless link $\overrightarrow{x_i x_j}$ as

$$P_{\mathrm{rx}}(x_i, x_j) = P \cdot g(x_i, x_j), \quad (1)$$

where $P$ is the transmit power, and $g(x_i, x_j)$ is the power gain of the link $\overrightarrow{x_i x_j}$. The gain $g(x_i, x_j)$ is considered constant (quasi-static) throughout the use of the communications channel, corresponding to channels with a large coherence time. Furthermore, the function $g$ is assumed to satisfy the following conditions, which are typically observed in practice: i) $g(x_i, x_j)$ depends on $x_i$ and $x_j$ only through the link length $|x_i - x_j|$;[1] ii) $g(r)$ is continuous and strictly decreasing with $r$; and iii) $\lim_{r \to \infty} g(r) = 0$.

### B. $i\mathcal{S}$-Graph

Consider a wireless network where the legitimate nodes and the potential eavesdroppers are randomly scattered in space, according to some point processes. The $i\mathcal{S}$-graph is a convenient representation of the links that can be established with information-theoretic security on such a network.

*Definition 2.1 ($i\mathcal{S}$-Graph [8]):* Let $\Pi_\ell = \{x_i\} \subset \mathbb{R}^d$ denote the set of legitimate nodes, and $\Pi_e = \{e_i\} \subset \mathbb{R}^d$ denote the set of eavesdroppers. The $i\mathcal{S}$-graph is the directed graph $G = \{\Pi_\ell, \mathcal{E}\}$ with vertex set $\Pi_\ell$ and edge set

$$\mathcal{E} = \{\overrightarrow{x_i x_j} : \mathcal{R}_{\mathrm{s}}(x_i, x_j) > \varrho\}, \quad (2)$$

where $\varrho$ is a threshold representing the prescribed infimum secrecy rate for each communication link; and $\mathcal{R}_{\mathrm{s}}(x_i, x_j)$ is the *maximum secrecy rate* (MSR) of the link $\overrightarrow{x_i x_j}$, given by

$$\mathcal{R}_{\mathrm{s}}(x_i, x_j) = \left[\log_2\left(1 + \frac{P_{\mathrm{rx}}(x_i, x_j)}{\sigma_\ell^2}\right) - \log_2\left(1 + \frac{P_{\mathrm{rx}}(x_i, e^*)}{\sigma_e^2}\right)\right]^+ \quad (3)$$

---
[1]With abuse of notation, we can write $g(r) \triangleq g(x_i, x_j)|_{|x_i - x_j| \to r}$.

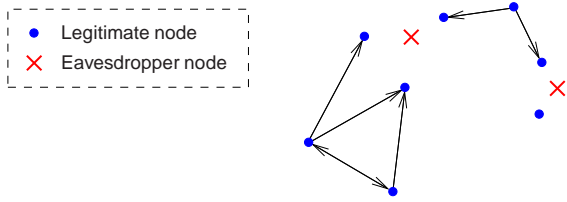

Figure 1. Example of an $i\mathcal{S}$-graph on $\mathbb{R}^2$.

in bits per complex dimension, where $[x]^+ = \max\{x,0\}$; $\sigma_\ell^2, \sigma_e^2$ are the noise powers of the legitimate users and eavesdroppers, respectively; and $e^* = \underset{e_k \in \Pi_e}{\operatorname{argmax}} P_{\text{rx}}(x_i, e_k)$.[2]

In the remainder of the paper, we consider that the noise powers of the legitimate users and eavesdroppers are equal, i.e., $\sigma_\ell^2 = \sigma_e^2 = \sigma^2$. In such case, we can combine (1)-(3) to obtain the following edge set

$$\mathcal{E} = \left\{ \overrightarrow{x_i x_j} : g(|x_i - x_j|) > 2^\varrho g(|x_i - e^*|) + \frac{\sigma^2}{P}(2^\varrho - 1) \right\}, \quad (4)$$

where $e^* = \underset{e_k \in \Pi_e}{\operatorname{argmin}} |x_i - e_k|$ is the eavesdropper closest to the transmitter $x_i$. The particular case of $\varrho = 0$ corresponds to considering the *existence* of secure links, in the sense that an edge $\overrightarrow{x_i x_j}$ is present iff $\mathcal{R}_s(x_i, x_j) > 0$. In such case, the edge set in (4) simplifies to

$$\mathcal{E} = \left\{ \overrightarrow{x_i x_j} : |x_i - x_j| < |x_i - e^*|, \quad e^* = \underset{e_k \in \Pi_e}{\operatorname{argmin}} |x_i - e_k| \right\}.$$

Fig. 1 shows an example of such an $i\mathcal{S}$-graph on $\mathbb{R}^2$.

The spatial location of the legitimate and eavesdropper nodes can be modeled either deterministically or stochastically. In many cases, the node positions are unknown to the network designer a priori, so they may be treated as uniformly random according to a Poisson point process [12]–[14].

*Definition 2.2 (Poisson $i\mathcal{S}$-Graph):* The *Poisson $i\mathcal{S}$-graph* is an $i\mathcal{S}$-graph where $\Pi_\ell, \Pi_e \subset \mathbb{R}^d$ are mutually independent, homogeneous Poisson point processes with densities $\lambda_\ell$ and $\lambda_e$, respectively.

In the remainder of the paper (unless otherwise indicated), we focus on Poisson $i\mathcal{S}$-graphs in $\mathbb{R}^2$.

## III. DEFINITIONS

*Graphs:* We use $G = \{\Pi_\ell, \mathcal{E}\}$ to denote the (directed) $i\mathcal{S}$-graph with vertex set $\Pi_\ell$ and edge set given in (2). In addition, we define two undirected graphs: the *weak $i\mathcal{S}$-graph* $G^{\text{weak}} = \{\Pi_\ell, \mathcal{E}^{\text{weak}}\}$, where

$$\mathcal{E}^{\text{weak}} = \{\overline{x_i x_j} : \mathcal{R}_s(x_i, x_j) > \varrho \vee \mathcal{R}_s(x_j, x_i) > \varrho\},$$

and the *strong $i\mathcal{S}$-graph* $G^{\text{strong}} = \{\Pi_\ell, \mathcal{E}^{\text{strong}}\}$, where

$$\mathcal{E}^{\text{strong}} = \{\overline{x_i x_j} : \mathcal{R}_s(x_i, x_j) > \varrho \wedge \mathcal{R}_s(x_j, x_i) > \varrho\}.$$

[2]This definition uses *strong secrecy* as the condition for information-theoretic security. See [8], [11] for more details.

*Graph Components:* We use the notation $x \overset{G}{\to} y$ to represent a path from node $x$ to node $y$ in a directed graph $G$, and $x \overset{G^*}{-} y$ to represent a path between node $x$ and node $y$ in an undirected graph $G^*$. We define four components:

$$\mathcal{K}^{\text{out}}(x) \triangleq \{y \in \Pi_\ell : \exists\, x \overset{G}{\to} y\}, \quad (5)$$

$$\mathcal{K}^{\text{in}}(x) \triangleq \{y \in \Pi_\ell : \exists\, y \overset{G}{\to} x\}, \quad (6)$$

$$\mathcal{K}^{\text{weak}}(x) \triangleq \{y \in \Pi_\ell : \exists\, x \overset{G^{\text{weak}}}{-} y\}, \quad (7)$$

$$\mathcal{K}^{\text{strong}}(x) \triangleq \{y \in \Pi_\ell : \exists\, x \overset{G^{\text{strong}}}{-} y\}. \quad (8)$$

*Percolation Probabilities:* To study percolation in the $i\mathcal{S}$-graph, it is useful to define percolation probabilities associated with the four graph components. Specifically, let $p_\infty^{\text{out}}$, $p_\infty^{\text{in}}$, $p_\infty^{\text{weak}}$, and $p_\infty^{\text{strong}}$ respectively be the probabilities that the in, out, weak, and strong components containing node $x = 0$ have an infinite number of nodes, i.e.,[3]

$$p_\infty^\diamond(\lambda_\ell, \lambda_e, \varrho) \triangleq \mathbb{P}\{|\mathcal{K}^\diamond(0)| = \infty\}$$

for $\diamond \in \{\text{out}, \text{in}, \text{weak}, \text{strong}\}$.[4] Our goal is to study the properties and behavior of these percolation probabilities.

## IV. MAIN RESULT

Typically, a continuum percolation model consists of an underlying point process defined on the infinite plane, and a rule that describes how connections are established between the nodes [5]. A main property of all percolation models is that they exhibit a *phase transition* as some continuous parameter is varied. If this parameter is the density $\lambda$ of nodes, then the phase transition occurs at some *critical density* $\lambda_c$. When $\lambda < \lambda_c$, denoted as the *subcritical phase*, all the clusters are a.s. bounded.[5] When $\lambda > \lambda_c$, denoted as the *supercritical phase*, the graph exhibits a.s. an unbounded cluster of nodes, or in other words, the graph *percolates*.

In this section, we aim to determine if percolation in the $i\mathcal{S}$-graph is possible, and if so, for which combinations of system parameters $(\lambda_\ell, \lambda_e, \varrho)$ does it occur. The mathematical characterization of the $i\mathcal{S}$-graph presents two challenges: i) the $i\mathcal{S}$-graph is a directed graph, which leads to the study of *directed percolation*; and ii) the $i\mathcal{S}$-graph exhibits dependencies between the state of different edges, which leads to the study of *dependent percolation*. The result is given by the following main theorem.

*Theorem 4.1 (Phase Transition in the $i\mathcal{S}$-Graph):* For any $\lambda_e > 0$ and $\varrho$ satisfying

$$0 \le \varrho < \varrho_{\max} \triangleq \log_2\left(1 + \frac{P \cdot g(0)}{\sigma^2}\right), \quad (9)$$

there exist critical densities $\lambda_c^{\text{out}}, \lambda_c^{\text{in}}, \lambda_c^{\text{weak}}, \lambda_c^{\text{strong}}$ satisfying

$$0 < \lambda_c^{\text{weak}} \le \lambda_c^{\text{out}} \le \lambda_c^{\text{strong}} < \infty \quad (10)$$

$$0 < \lambda_c^{\text{weak}} \le \lambda_c^{\text{in}} \le \lambda_c^{\text{strong}} < \infty \quad (11)$$

[3]We condition on the event that a legitimate node is located at $x = 0$. According to Slivnyak's theorem [15, Sec. 4.4], apart from the given point at $x = 0$, the probabilistic structure of the conditioned process is identical to that of the original process.

[4]Except where otherwise indicated, we use the symbol $\diamond$ to represent the out, in, weak, or strong component.

[5]We say that an event occurs "almost surely" (a.s.) if its probability is equal to one.

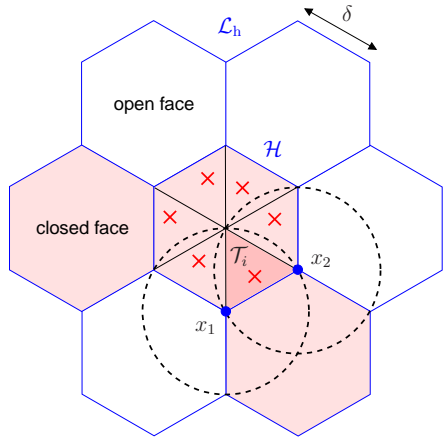 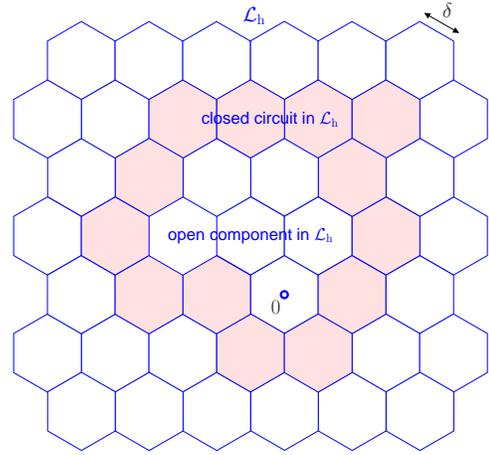

(a) Conditions for a face $\mathcal{H}$ in $\mathcal{L}_\text{h}$ to be closed, according to Definition 5.1.

(b) A finite open component at the origin, surrounded by a closed circuit.

Figure 2. Auxiliary diagrams for proving Lemma 4.1.

such that

$$p_\infty^\diamond = 0, \quad \text{for } \lambda_\ell < \lambda_\text{c}^\diamond, \quad (12)$$
$$p_\infty^\diamond > 0, \quad \text{for } \lambda_\ell > \lambda_\text{c}^\diamond, \quad (13)$$

for any $\diamond \in \{\text{out}, \text{in}, \text{weak}, \text{strong}\}$. Conversely, if $\varrho > \varrho_\text{max}$, then $p_\infty^\diamond = 0$ for any $\lambda_\ell, \lambda_\text{e}$.

To prove the theorem, we introduce the following three lemmas.

*Lemma 4.1:* For any $\lambda_\text{e} > 0$ and $\varrho$ satisfying (9), there exists an $\epsilon > 0$ such that $p_\infty^\text{weak}(\lambda_\ell) = 0$ for all $\lambda_\ell < \epsilon$.

*Proof:* Postponed to Section V of the present paper. □

*Lemma 4.2:* For any $\lambda_\text{e} > 0$ and $\varrho$ satisfying (9), there exists a $\zeta < \infty$ such that $p_\infty^\text{strong}(\lambda_\ell) > 0$ for all $\lambda_\ell > \zeta$.

*Proof:* Postponed to Section VI of the present paper. □

*Lemma 4.3:* For any $\lambda_\text{e} > 0$ and $\varrho \geq 0$, the percolation probability $p_\infty^\diamond(\lambda_\ell)$ is a non-decreasing function of $\lambda_\ell$.

*Proof:* This follows directly from a coupling argument. The details can be found in [16]. □

With these lemmas we are now in condition to prove Theorem 4.1.

*Proof of Theorem 4.1:* We first show that if $\varrho > \varrho_\text{max}$, then $p_\infty^\diamond = 0$. The MSR $\mathcal{R}_\text{s}$ of a link $\overrightarrow{x_i x_j}$, given in (3), is upper bounded by the channel capacity $\mathcal{R}$ of a link with zero length, i.e., $\mathcal{R}_\text{s}(x_i, x_j) \leq \mathcal{R}(x_i, x_i) = \log_2\left(1 + \frac{P \cdot g(0)}{\sigma^2}\right)$. If the threshold $\varrho$ is set such that $\varrho > \varrho_\text{max}$, the condition $\mathcal{R}_\text{s}(x_i, x_j) > \varrho$ in (2) for the existence of the edge $\overrightarrow{x_i x_j}$ is never satisfied by any $x_i, x_j$. Thus, the $i\mathcal{S}$-graph $G$ has no edges and cannot percolate. We now consider the case of $0 \leq \varrho < \varrho_\text{max}$. From the definitions in (5)-(8), we have $\mathcal{K}^\text{strong}(0) \subseteq \mathcal{K}^\text{out}(0) \subseteq \mathcal{K}^\text{weak}(0)$ and $\mathcal{K}^\text{strong}(0) \subseteq \mathcal{K}^\text{in}(0) \subseteq \mathcal{K}^\text{weak}(0)$, and therefore $p_\infty^\text{strong} \leq p_\infty^\text{out} \leq p_\infty^\text{weak}$ and $p_\infty^\text{strong} \leq p_\infty^\text{in} \leq p_\infty^\text{weak}$. Then, Lemmas 4.1, 4.2, and 4.3 imply that each curve $p_\infty^\diamond(\lambda_\ell)$ departs from the zero value at some critical density $\lambda_\text{c}^\diamond$, as expressed by (12) and (13). Furthermore, these critical densities are nontrivial in the sense that $0 < \lambda_\text{c}^\diamond < \infty$. The ordering of the critical densities in (10) and (11) follows from similar coupling arguments. □

## V. PROOF OF LEMMA 4.1

In this proof, it is sufficient to show that $G^\text{weak}$ does not percolate for sufficiently small $\lambda_\ell$ *in the case of $\varrho = 0$*, since for larger $\varrho$ the connectivity is worse and $G^\text{weak}$ certainly does not percolate either.[6]

### A. Mapping on a Lattice

We start with some definitions. Let $\mathcal{L}_\text{h}$ be an hexagonal lattice as depicted in Fig. 2(a), where each face is a regular hexagon with side length $\delta$. Each face has a *state*, which can be either *open* or *closed*. A set of faces (e.g., a path or a circuit) in $\mathcal{L}_\text{h}$ is said to be open iff all the faces that form the set are open. We now specify the state of a face based on how the processes $\Pi_\ell$ and $\Pi_\text{e}$ behave inside that face.

*Definition 5.1 (Closed Face in $\mathcal{L}_\text{h}$):* A face $\mathcal{H}$ in $\mathcal{L}_\text{h}$ is said to be *closed* iff all the following conditions are met:
1) Each of the 6 equilateral triangles $\{\mathcal{T}_i\}_{i=1}^6$ that compose the hexagon $\mathcal{H}$ has at least one eavesdropper.
2) The hexagon $\mathcal{H}$ is free of legitimate nodes.

The above definition was chosen such that discrete face percolation in $\mathcal{L}_\text{h}$ can be tied to continuum percolation in $G^\text{weak}$, as given by the following proposition.

*Proposition 5.1 (Circuit Coupling):* If there exists a closed circuit in $\mathcal{L}_\text{h}$ surrounding the origin, then the component $\mathcal{K}^\text{weak}(0)$ is finite.

*Proof:* Assume there is a closed circuit $\mathcal{C}$ in $\mathcal{L}_\text{h}$ surrounding the origin, as depicted in Fig. 2(b). This implies that the open component in $\mathcal{L}_\text{h}$ containing 0, denoted by $\mathcal{K}^{\mathcal{L}_\text{h}}(0)$, must be finite. Since the area of the region $\mathcal{K}^{\mathcal{L}_\text{h}}(0)$ is finite, the continuous graph $G^\text{weak}$ has a finite number of vertices falling inside this region. Thus, to prove that $\mathcal{K}^\text{weak}(0)$ is finite, we just need to show that no edges of $G^\text{weak}$ cross the circuit $\mathcal{C}$. Consider Fig. 2(a), and suppose that the shaded faces are part of the closed circuit $\mathcal{C}$. Let $x_1, x_2$ be two legitimate nodes such that $x_1$ falls on an open face on the inner side of $\mathcal{C}$, while $x_2$ falls on the outer side of $\mathcal{C}$. Clearly, the most favorable situation for $x_1, x_2$ being able to establish an edge across $\mathcal{C}$ is the one depicted in Fig. 2(a). The edge $\overline{x_1 x_2}$ exists in $G^\text{weak}$ iff either $\mathcal{B}_{x_1}(\delta)$ or $\mathcal{B}_{x_2}(\delta)$ are free of eavesdroppers.[7] This condition

---

[6] A simple coupling argument shows that the percolation probabilities $p_\infty^\diamond(\lambda_\ell, \lambda_\text{e}, \varrho)$ are non-increasing functions of $\varrho$.

[7] We use $\mathcal{B}_x(\rho) \triangleq \{y \in \mathbb{R}^2 : |y - x| \leq \rho\}$ to denote the closed two-dimensional ball centered at point $x$, with radius $\rho$.

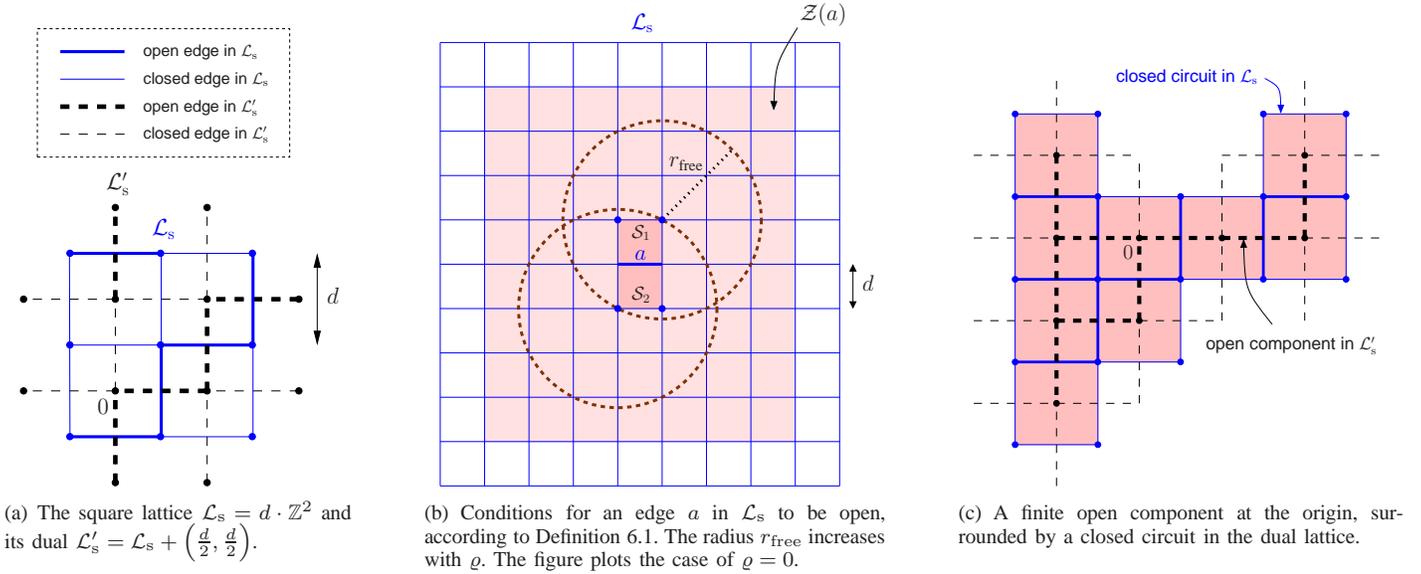

(a) The square lattice $\mathcal{L}_s = d \cdot \mathbb{Z}^2$ and its dual $\mathcal{L}'_s = \mathcal{L}_s + \left(\frac{d}{2}, \frac{d}{2}\right)$.

(b) Conditions for an edge $a$ in $\mathcal{L}_s$ to be open, according to Definition 6.1. The radius $r_{\text{free}}$ increases with $\varrho$. The figure plots the case of $\varrho = 0$.

(c) A finite open component at the origin, surrounded by a closed circuit in the dual lattice.

Figure 3. Auxiliary diagrams for proving Lemma 4.2.

does not hold, since Definition 5.1 guarantees that at least one eavesdropper is located inside the triangle $\mathcal{T}_i \subset \mathcal{B}_{x_1}(\delta) \cap \mathcal{B}_{x_2}(\delta)$. Thus, no edges of $G^{\text{weak}}$ cross the circuit $\mathcal{C}$, which implies that $\mathcal{K}^{\text{weak}}(0)$ has finite size. □

### B. Discrete Percolation

Having performed an appropriate mapping from a continuous to a discrete model, we now analyze discrete face percolation in $\mathcal{L}_h$.

*Proposition 5.2 (Discrete Percolation in $\mathcal{L}_h$):* If the parameters $\lambda_\ell, \lambda_e, \delta$ satisfy

$$\left(1 - e^{-\lambda_e \frac{\sqrt{3}}{4}\delta^2}\right)^6 \cdot e^{-\lambda_\ell \frac{3\sqrt{3}}{2}\delta^2} > \frac{1}{2}, \quad (14)$$

then the origin is a.s. surrounded by a closed circuit in $\mathcal{L}_h$.

*Proof:* The model introduced in Section V-A can be seen as a face percolation model on the hexagonal lattice $\mathcal{L}_h$, where each face is closed independently of other faces with probability

$$p \triangleq \mathbb{P}\{\text{face } \mathcal{H} \text{ of } \mathcal{L}_h \text{ is closed}\}$$
$$= \mathbb{P}\left\{\left(\bigwedge_{i=1}^{6} \Pi_e\{\mathcal{T}_i\} \geq 1\right) \wedge \Pi_\ell\{\mathcal{H}\} = 0\right\}$$
$$= \left(1 - e^{-\lambda_e \frac{\sqrt{3}}{4}\delta^2}\right)^6 \cdot e^{-\lambda_\ell \frac{3\sqrt{3}}{2}\delta^2}. \quad (15)$$

Face percolation on the hexagonal lattice can be equivalently described as site percolation on the triangular lattice. In particular, recall that if

$$\mathbb{P}\{\mathcal{H} \text{ is open}\} < \frac{1}{2}, \quad (16)$$

then the *open* component in $\mathcal{L}_h$ containing the origin is a.s. finite [17, Ch. 5, Thm. 8], and so the origin is a.s. surrounded by a *closed* circuit in $\mathcal{L}_h$. Combining (15) and (16), we obtain (14). □

We now use the propositions to finalize the proof of Lemma 4.1, whereby $p_\infty^{\text{weak}}(\lambda_\ell) = 0$ for sufficiently small (but nonzero) $\lambda_\ell$.

*Proof of Lemma 4.1:* For any fixed $\lambda_e$, it is easy to see that condition (14) can always be met by making the hexagon side $\delta$ large enough, and the density $\lambda_\ell$ small enough (but nonzero). For any such choice of parameters $\lambda_\ell, \lambda_e, \delta$ satisfying (14), the origin is a.s. surrounded by a closed circuit in $\mathcal{L}_h$ (by Proposition 5.2), and the component $\mathcal{K}^{\text{weak}}(0)$ is a.s. finite (by Proposition 5.1), i.e., $p_\infty^{\text{weak}}(\lambda_\ell) = 0$. □

## VI. Proof of Lemma 4.2

### A. Mapping on a Lattice

We start with some definitions. Let $\mathcal{L}_s \triangleq d \cdot \mathbb{Z}^2$ be a square lattice with edge length $d$. Let $\mathcal{L}'_s \triangleq \mathcal{L}_s + \left(\frac{d}{2}, \frac{d}{2}\right)$ be the dual lattice of $\mathcal{L}_s$, depicted in Fig. 3(a). We make the origin of the coordinate system coincide with a vertex of $\mathcal{L}'_s$. Each edge has a *state*, which can be either *open* or *closed*. We declare an edge $a'$ in $\mathcal{L}'_s$ to be open iff its dual edge $a$ in $\mathcal{L}_s$ is open.

We now specify the state of an edge based on how the processes $\Pi_\ell$ and $\Pi_e$ behave in the neighborhood of that edge. Consider Fig. 3(b), where $a$ denotes an edge in $\mathcal{L}_s$, and $\mathcal{S}_1(a)$ and $\mathcal{S}_2(a)$ denote the two squares adjacent to $a$. Let $\{v_i\}_{i=1}^4$ denote the four vertices of the rectangle $\mathcal{S}_1(a) \cup \mathcal{S}_2(a)$. We then have the following definition.

*Definition 6.1 (Open Edge in $\mathcal{L}_s$):* An edge $a$ in $\mathcal{L}_s$ is said to be *open* iff all the following conditions are met:
1) Each square $\mathcal{S}_1(a)$ and $\mathcal{S}_2(a)$ adjacent to $a$ has at least one legitimate node.
2) The region $\mathcal{Z}(a)$ is free of eavesdroppers, where $\mathcal{Z}(a)$ is smallest rectangle such that $\bigcup_{i=1}^{4} \mathcal{B}_{v_i}(r_{\text{free}}) \subset \mathcal{Z}(a)$ with[8]

$$r_{\text{free}} \triangleq g^{-1}\left(2^{-\varrho}g(\sqrt{5}d) - \frac{\sigma^2}{P}(1 - 2^{-\varrho})\right). \quad (17)$$

The above definition was chosen such that discrete edge percolation in $\mathcal{L}'_s$ can be tied to continuum percolation in $G^{\text{strong}}$, as given by the following two propositions.

---

[8]To ensure that $r_{\text{free}}$ in (17) is well-defined, in the rest of the paper we assume that $d$ is chosen such that $d < \frac{1}{\sqrt{5}}g^{-1}\left(\frac{\sigma^2}{P}(2^\varrho - 1)\right)$.

*Proposition 6.1 (Two-Square Coupling):* If $a$ is an open edge in $\mathcal{L}_s$, then all legitimate nodes inside $\mathcal{S}_1(a) \cup \mathcal{S}_2(a)$ form a single connected component in $G^{\text{strong}}$.

*Proof:* If two legitimate nodes $x_1, x_2$ are to be placed inside the region $\mathcal{S}_1(a) \cup \mathcal{S}_2(a)$, the most unfavorable configuration in terms of MSR is the one where $x_1$ and $x_2$ are on opposite corners of the rectangle $\mathcal{S}_1(a) \cup \mathcal{S}_2(a)$ and thus $|x_1 - x_2| = \sqrt{5}d$. In such configuration, we see from (4) that the edge $\overrightarrow{x_1 x_2}$ exists in $G$ iff $|x_1 - e^*| > g^{-1}\left(2^{-\varrho}g(\sqrt{5}d) - \frac{\sigma^2}{P}(1 - 2^{-\varrho})\right) \triangleq r_{\text{free}}$, where $e^*$ is the eavesdropper closest to $x_1$. As a result, the edge $\overline{x_1 x_2}$ exists in $G^{\text{strong}}$ iff both $\mathcal{B}_{x_1}(r_{\text{free}})$ and $\mathcal{B}_{x_2}(r_{\text{free}})$ are free of eavesdroppers. Now, if $\mathcal{Z}(a)$ is the smallest rectangle containing the region $\bigcup_{i=1}^{4} \mathcal{B}_{v_i}(r_{\text{free}})$, the condition $\Pi_e\{\mathcal{Z}(a)\} = 0$ ensures the edge $\overline{x_i x_j}$ exists in $G^{\text{strong}}$ for any $x_i, x_j \in \mathcal{S}_1(a) \cup \mathcal{S}_2(a)$. Thus, all legitimate nodes inside $\mathcal{S}_1(a) \cup \mathcal{S}_2(a)$ form a single connected component in $G^{\text{strong}}$. □

*Proposition 6.2 (Component Coupling):* If the open component in $\mathcal{L}'_s$ containing the origin is infinite, then the component $\mathcal{K}^{\text{strong}}(0)$ is also infinite.

*Proof:* Consider Fig. 3(c). Let $\mathcal{P} = \{a'_i\}$ denote a path of open edges $\{a'_i\}$ in $\mathcal{L}'_s$. By the definition of dual lattice, the path $\mathcal{P}$ uniquely defines a sequence $\mathcal{S} = \{\mathcal{S}_i\}$ of *adjacent* squares in $\mathcal{L}_s$, separated by open edges $\{a_i\}$ (the duals of $\{a'_i\}$). Then, each square in $\mathcal{S}$ has at least one legitimate node (by Definition 6.1), and all legitimate nodes falling inside the region associated with $\mathcal{S}$ form a single connected component in $G^{\text{strong}}$ (by Proposition 6.1). Now, let $\mathcal{K}^{\mathcal{L}'_s}(0)$ denote the open component in $\mathcal{L}'_s$ containing 0. Because of the argument just presented, we have $|\mathcal{K}^{\mathcal{L}'_s}(0)| \leq |\mathcal{K}^{\text{strong}}(0)|$. Thus, if $|\mathcal{K}^{\mathcal{L}'_s}(0)| = \infty$, then $|\mathcal{K}^{\text{strong}}(0)| = \infty$. □

### B. Discrete Percolation

Having performed an appropriate mapping from a continuous to a discrete model, we now analyze discrete edge percolation in $\mathcal{L}'_s$. Let $N_s$ be the number of squares that compose the rectangle $\mathcal{Z}(a)$ introduced in Definition 6.1. We first study the behavior of paths in $\mathcal{L}_s$ with the following proposition.

*Proposition 6.3 (Geometric Bound):* The probability that a given path in $\mathcal{L}_s$ with length $n$ is closed is bounded by

$$\mathbb{P}\{\text{path in } \mathcal{L}_s \text{ with length } n \text{ is closed}\} \leq q^{n/N_e}, \quad (18)$$

where $N_e$ is a finite integer and

$$q = 1 - (1 - e^{-\lambda_\ell d^2})^2 \cdot e^{-\lambda_e N_s d^2} \quad (19)$$

is the probability that an edge in $\mathcal{L}_s$ is closed.

*Proof: (outline)* Using Definition 6.1, we can write

$$q \triangleq \mathbb{P}\{\text{edge } a \text{ in } \mathcal{L}_s \text{ is closed}\}$$
$$= 1 - \mathbb{P}\{\Pi_\ell\{\mathcal{S}_1(a)\} \geq 1 \wedge \Pi_\ell\{\mathcal{S}_2(a)\} \geq 1 \wedge \Pi_e\{\mathcal{Z}(a)\} = 0\}$$
$$= 1 - (1 - e^{-\lambda_\ell d^2})^2 \cdot e^{-\lambda_e N_s d^2}.$$

Now, let $\mathcal{P} = \{a_i\}_{i=1}^{n}$ denote a path in $\mathcal{L}_s$ with length $n$ and edges $\{a_i\}$. Even though the edges $\{a_i\}$ do not all have independent states (in which case we would have $\mathbb{P}\{\mathcal{P} \text{ is closed}\} = q^n$), it is possible to show that $\mathbb{P}\{\mathcal{P} \text{ is closed}\} \leq q^{n/N_e}$ for a finite integer $N_e$, by finding a subset $\mathcal{Q} \subseteq \mathcal{P}$ of edges with independent states (see [16] for details). □

Having obtained a geometric bound on the probability of a path of length $n$ being closed, we can now use a Peierls argument to study the existence of an infinite component.[9]

*Proposition 6.4 (Discrete Percolation in $\mathcal{L}'_s$):* If the probability $q$ satisfies

$$q < \left(\frac{11 - 2\sqrt{10}}{27}\right)^{N_e}, \quad (20)$$

then

$$\mathbb{P}\{\text{open component in } \mathcal{L}'_s \text{ containing 0 is infinite}\} > 0. \quad (21)$$

*Proof:* We start with the key observation that the open component in $\mathcal{L}'_s$ containing 0 is *finite* iff there is a closed circuit in $\mathcal{L}_s$ surrounding 0, as illustrated in Fig. 3(c). Thus, the inequality in (21) is equivalent to $\mathbb{P}\{\exists \text{ closed circuit in } \mathcal{L}_s \text{ surrounding } 0\} < 1$. Let $\rho(n)$ denote the possible number of circuits of length $n$ in $\mathcal{L}_s$ surrounding 0 (a deterministic quantity). Let $\kappa(n)$ denote the number of *closed* circuits of length $n$ in $\mathcal{L}_s$ surrounding 0 (a random variable). From combinatorial arguments, it can be shown [19, Eq. (1.17)] that $\rho(n) \leq 4n3^{n-2}$. Then, for a fixed $n$,

$$\mathbb{P}\{\kappa(n) \geq 1\} \leq \rho(n) \cdot \mathbb{P}\{\text{path in } \mathcal{L}_s \text{ with length } n \text{ is closed}\}$$
$$\leq 4n3^{n-2}q^{n/N_e},$$

where we used the union bound and Proposition 6.3. Also,

$$\mathbb{P}\{\exists \text{ closed circuit in } \mathcal{L}_s \text{ surrounding } 0\}$$
$$= \mathbb{P}\{\kappa(n) \geq 1 \text{ for some } n\}$$
$$\leq \sum_{n=1}^{\infty} 4n3^{n-2}q^{n/N_e} = \frac{4q^{1/N_e}}{3(1 - 3q^{1/N_e})^2}, \quad (22)$$

for $q < \left(\frac{1}{3}\right)^{N_e}$. We see that if $q$ satisfies (20), then (22) is strictly less than one, and (21) follows. □

We now use the propositions to finalize the proof of Lemma 4.2, whereby $p_\infty^{\text{strong}}(\lambda_\ell) > 0$ for sufficiently large (but finite) $\lambda_\ell$.

*Proof of Lemma 4.2:* For any fixed $\lambda_e$, it is easy to see the probability $q$ in (19) can be made small enough to satisfy condition (20), by making the edge length $d$ sufficiently small, and the density $\lambda_\ell$ sufficiently large (but finite). For any such choice of parameters $\lambda_\ell, \lambda_e, d$ satisfying (20), the open component in $\mathcal{L}'_s$ containing 0 is infinite with positive probability (by Proposition 6.4), and the component $\mathcal{K}^{\text{strong}}(0)$ is also infinite with positive probability (by Proposition 6.2), i.e., $p_\infty^{\text{strong}}(\lambda_\ell) > 0$. □

## VII. SIMULATION RESULTS

In this section, we obtain additional insights about percolation in the $i\mathcal{S}$-graph via Monte Carlo simulation. Figure 4 shows the percolation probabilities for the weak and strong components of the $i\mathcal{S}$-graph, versus the density $\lambda_\ell$ of legitimate nodes. As predicted by Theorem 4.1, the figure suggests that these components experience a phase transition as $\lambda_\ell$ is increased. In particular, $\lambda_c^{\text{weak}} \approx 3.4 \, \text{m}^{-2}$ and $\lambda_c^{\text{strong}} \approx 6.2 \, \text{m}^{-2}$, for the case of $\lambda_e = 1 \, \text{m}^{-2}$ and $\varrho = 0$. Operationally, this means that if long-range bidirectional secure communication is desired in a wireless network, the density of legitimate nodes must be at least 6.2 times that of the eavesdroppers. In practice, the density of legitimate nodes must be even larger, because a secrecy requirement greater than $\varrho = 0$ is typically required. This dependence on $\varrho$ is illustrated in Figure 5. In practice, it might also be of interest to increase $\lambda_\ell$ fairly beyond the critical density, since this leads to an increased average fraction $p_\infty^\diamond$ of nodes which belong to the infinite component, thus improving secure connectivity.

---
[9]A "Peierls argument", so-named in honour of Rudolf Peierls and his 1936 article on the Ising model [18], refers to an approach based on enumeration.

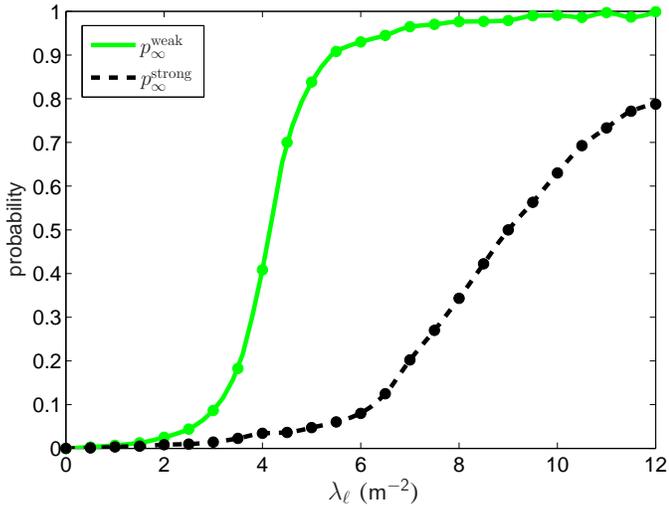

Figure 4. Simulated percolation probabilities for the weak and strong components of the $i\mathcal{S}$-graph, versus the density $\lambda_\ell$ of legitimate nodes ($\lambda_e = 1\,\mathrm{m}^{-2}$, $\varrho = 0$).

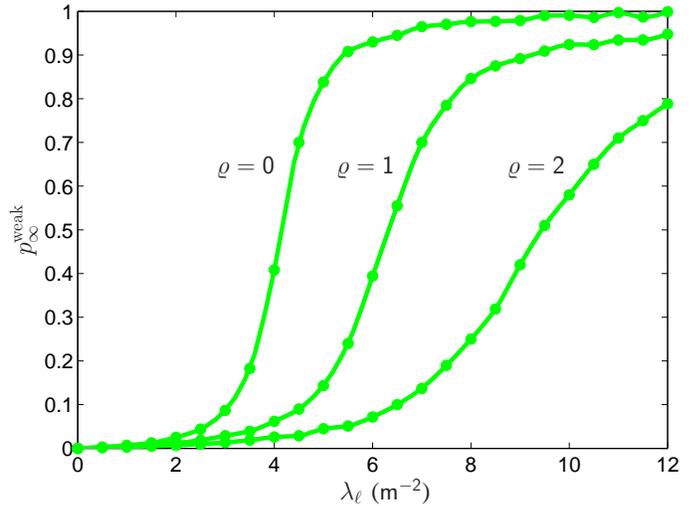

Figure 5. Effect of the secrecy rate threshold $\varrho$ on the percolation probability $p_\infty^{\mathrm{weak}}$ ($\lambda_e = 1\,\mathrm{m}^{-2}$, $g(r) = 1/r^4$, $P_\ell/\sigma^2 = 10$).

## VIII. Discussion and Conclusions

Theorem 4.1 shows that each of the four components of the $i\mathcal{S}$-graph (in, out, weak, and strong) experiences a phase transition at some nontrivial critical density $\lambda_c^\diamond$ of legitimate nodes. Operationally, this implies that long-range communication over multiple hops is still feasible when a secrecy constraint is present. We proved that percolation can occur for any prescribed secrecy threshold $\varrho$ satisfying $\varrho < \varrho_{\max} = \log_2\left(1 + \frac{P \cdot g(0)}{\sigma^2}\right)$, as long as the density of legitimate nodes is made large enough. This implies that for unbounded path loss models such as $g(r) = 1/r^\gamma$, percolation can occur for *any* arbitrarily large secrecy requirement $\varrho$, while for bounded models such as $g(r) = 1/(1 + r^\gamma)$, the desired $\varrho$ may be too high to allow percolation. Our results also show that as long as $\varrho < \varrho_{\max}$, percolation can be achieved even in cases where the eavesdroppers are arbitrarily dense, by making the density of legitimate nodes large enough. Using Monte Carlo simulation, we obtained numerical estimates for various critical densities. For example, when $\varrho = 0$ we estimated that if the density of eavesdroppers is larger than roughly $30\%$ that of the legitimate nodes, long-range communication in the weak $i\mathcal{S}$-graph is completely disrupted, in the sense that no infinite cluster arises. In the strong $i\mathcal{S}$-graph, we estimated this fraction to be about $16\%$. For a larger secrecy requirement $\varrho$, an even more modest fraction of attackers is enough to disrupt the network. We are hopeful that further efforts will lead to tight analytical bounds for these critical densities.